\documentstyle[aps]{revtex}
\begin{document}
\pagestyle{myheadings}
\markright{Submitted to Phys. Rev. Lett.}
\twocolumn
\draft
\title{Growth Kinetics in  Multicomponent Fluids}
\author{Shiyi Chen and Turab Lookman*}

\address{Theoretical Division and Center for Nonlinear Studies,\\
Los Alamos National Laboratory, Los Alamos NM 87545}

\date{\today}
\maketitle
\begin{abstract}
The hydrodynamic effects on the late stage
kinetics in spinodal decomposition of multicomponent fluids are examined using
a lattice Boltzmann scheme with stochastic fluctuations in the fluid and at the
interface. In two dimensions, the three and four component
immiscible fluid mixture
(with a $1024^2$ lattice) behaves like an off-critical binary
fluid with an estimated domain growth of $t^{0.4\pm.03}$ rather than $t^{1/3}$
as
previously predicted, showing the significant influence of
hydrodynamics. In three dimensions (with a $256^3$ lattice), we
estimate the growth as $t^{0.96\pm0.05}$ for both critical
and off-critical quenching, in agreement with phenomenological
theory.
\end{abstract}
\par It is well known that a binary fluid mixture undergoes phase
separation if rapidly quenched from a high temperature phase to a point in the
coexistence region. Moreover, when the sizes of the domains are much
larger than the interfacial thickness, there is only one dominant
length scale in the system\cite{rev}. It is accepted that the late-time
dynamics in a binary alloy or glass in which the order parameter is
conserved, follows a growth law of $R(t) \sim
t^{1/3}$, where $R(t)$ is the average size of the domains. This growth
law is  characteristic of the long range diffusion of particles
between domains and was first  predicted for off-critical quenches by
Lifshitz and Slyozov\cite{lif}.  Methods to
carry out simulations of phase segregating systems with hydrodynamic
interactions include molecular dynamics (MD), direct numerical
simulation of time-dependent Ginzburg-Landau equations, cell dynamical
systems and more recently, lattice gas \cite{ale} and lattice Boltzmann
models\cite{gru,alex}. Lattice
Boltzmann simulations have recently been successfully used to study spinodal
decomposition for critical quenches for a binary fluid in
 the presence\cite{gru}
and absence of porous media\cite{alex}. The various simulation techniques have
  tried to
address the question of the growth of single phase domains and the
scaling properties of the correlation or structure functions. The
underlying aim
 has been to find certain ``universality classes'' for first order
phase transitions in which the growth law is independent of the
details of the interactions, spatial dimension and the number of
components.

The aim of this work is to determine the growth law and scaling
properties in two and three dimensions for three and
four component fluids  with equal volume
fractions and compare the results to that for an off-critical binary
quench with unequal volume fractions. We obtain results in which the
hydrodynamic interactions are present as well as absent and show that
hydrodynamics is important.
In previous work on the LB method for studying phase segregation[4,5],
fluctuations due to correlations in the particles have been neglected.
 We include the effects of
fluctuations on domain growth in multicomponent systems by incorporating
stochastic fluctuations
in the fluid stress tensor using a scheme recently proposed by Ladd
\cite{ladd}
and fluctuations in the color gradient to perturb the interface.

 The influence of the number of components on domain growth and scaling has
been recently examined in two dimensions for a Potts model with two and three
components
using Monte Carlo \cite{jep} as
well as for a three component fluid using Molecular dynamics (MD)\cite{lar}
simulations. The Potts model does not
include hydrodynamics and therefore the result of $t^{1/3}$ for the
growth law for two and three components is not surprising. The
conclusion of the MD work for a three component fluid shows that
the growth exponent is also $1/3$.
Since a thorough MD study for fluids requires a large number of
particles and  has to be run for very long times, we have undertaken a
lattice Boltzmann approach while also incorporating fluctuations. Our
results demonstrate
 that for off-critical binary quenches and for  three and four
 component  fluids the growth exponent for late times scales
 as $t^{.4\pm.03}$ in two dimensions.
This is to be compared with $t^{1/3}$ if hydrodynamics were not
relevant  to phase segregation and domain growth in two  dimensions.
 Our three dimensional result is
the first simulation for an off-critical quench to verify the result of
Siggia\cite{siggia} that is based on a phenomenological model of droplet
coalescence.
 We also determine that dynamical scaling, which is a consequence of
the existence of one dominant length scale, is valid in two  and three
dimensions. The multicomponent fluids show the same scaling behaviour as an
off-critical binary quench.

Lattice gas and lattice Boltzmann (LB) methods can be used
 to study hydrodynamic phase segregation using
parallel computing techniques \cite{doolen}. They
simulate fluid properties, phase segregation and the interface
dynamics simultaneously and allow complex boundaries to be handled
easily. The methods have been described as providing the most
promising tools to study flow through porous media \cite{sah}.
 Unlike methods such as the use of the Langevin
equation\cite{far}  that
are based on a phenomenological model of fluid behaviour
 and are computationally intensive, LB
methods simulate hydrodynamic phase segregation in a natural way
without the introduction of {\em ad hoc} relations between the order
parameter fluctuations and the fluid dynamics. The lattice Boltzmann
method is a discrete, in space and time, microscopic kinetic equation
description for the evolution of the particle distribution function of
a fluid. Point particles move along the links of a lattice (hexagonal
in two dimensions), obey certain collision rules, and macroscopically
mimic the Navier-Stokes equations in certain limits.
 The LB two-phase model we used is
a modified version of the immiscible fluid model
proposed by Grunau {\em et al.}\cite{gru1} that is based on the original model
introduced by Gunstensen {\em et al.}\cite{gun}. The multicomponent
LB model in this paper is a extension of the lattice gas\cite{gun1} and lattice
Boltzmann \cite{gun2}
models by Gunstensen and Rothman.

The local order parameter is defined as  $\psi({\bf x},t)=\sum_{i=0}^{N}
(f_{i}^{1}({\bf x},t) - \sum_{k=2}^{n}f_{i}^{k}({\bf x},t) ), $ where
$f_{i}^{k}({\bf x},t)$
is the distribution function for the $k-th$ component $(k > 2)$ of the fluid
mixture
 at site ${\bf x}$ and time $t$ moving
along the link in the direction $i$. $n$ is the number of components and $N$ is
the number of velocity directions. Also,
$f_{i}({\bf x},t)=\sum_{k=1}^{n}f_{i}^{k}({\bf x},t)
$ is the distribution function for the total
fluid, where $i=0,1,\cdot\cdot\cdot,6$ represent the velocity directions at
each
site of a hexagonal lattice. The state $i=0$ corresponds to a portion
of the fluid at rest.  The LB equation for $f_{i}({\bf x},t)$ can be written
as
$f_{i}^{k}({\bf x} + {\bf e}_{i},t+1)=f_{i}^{k}({\bf x},t) +
\Omega_{i}^{k},$ where
{k} denotes the fluid component, and $(\Omega_{i}^{k})=
(\Omega_{i}^{k})^{c} + (\Omega_{i}^{k})^{p}$ is the collision
operator consisting of a term representing the rate of change of
$f_{i}^{k}$ due to collisions and a term representing the color
perturbation. The vectors ${\bf e}_{i}$ are the velocity vectors along the
links of a hexagonal lattice. The form of $(\Omega_{i}^{k})^{c}$ is
chosen to have a single time relaxation with
$(\Omega_{i}^{k})^{c}=-{1/{\tau}}(f_{i}^{(k)}-f_{i}^{(eq)})$,
where $\tau$ is the characteristic relaxation time and
$f_{i}^{(eq)}$ is the local equilibrium distribution\cite{chen}. The surface
tension inducing perturbation $(\Omega_{i}^{k})^{p}$ and the
recoloring
procedure are chosen appropriately  so that Laplaces's law holds for
the model \cite{gru1}.

The local color gradient ${\bf G}({\bf x})$ is defined by
${\bf G}({\bf x}) =\sum_{i=1}^{N}{\bf e}_i{\psi({\bf x} + {\bf e}_i)},$
and the surface color pertubation that is added to
segregate and stabilize the interface is
$(\Omega_{i}^{k})^{p} = A\mid {\bf G} \mid cos2(\theta_{i}-\theta_{G})$,
where $\theta_i$ is the angle of lattice direction $i$ and $\theta_G
=\arctan(G_y/G_x)$ is the angle of the local color gradient. The
surface tension is  proportional to  $\sim A \tau \rho$. Here
 $\rho = \sum_{i,k} f_i^{k}$ is the local
particle density.  In the original models\cite{gun,gru1}, the recoloring
step  always makes
the local color gradient to lie along the direction perpendicular to the
interface. To mimic the temperature effect on the interface, a noise is
introduced to perturb
the local color gradient direction $\theta_G$, by assuming that the angles are
distributed
according to a Gaussian distribution about $\theta_{G}$.
The variance of the angle distribution depends on some local
temperature.

A cause for concern about the use of the LB method for studying
spinodal decomposition has been the lack of statistical
correlations in the particles.
Previous simulations show that droplets form and collapse
due to an initial random concentration field. It is believed that the
spinodal decomposition process is strongly correlated with noise in
the system. It may be argued that the initial fluctuations may not be
equivalent to the natural noise due to the temperature of the system.
 In order to study the effects of fluctuations on domain
growth, we have incorporated stochastic fluctuations in the fluid
stress tensor according to a scheme proposed by Ladd\cite{ladd}.
The basic idea
is that, on length scales and time scales intermediate between
molecular and hydrodynamic, thermally-induced fluctuations can be
reduced to random fluctuations in the fluxes of conserved variables,
for instance, the stress tensor. It is thus plausible that in an LB
simulation, molecular fluctuations can be modeled realistically on
intermediate scales, although the microscopic interactions are
different from the real dynamics obtained from, for example,  an MD
simulation. To incorporate this effect, a stochastic term,
$f^{\prime}_{i}(r,t)$, representing the thermally induced fluctuations
in the stress tensor, is added to the time evolution of the density
distribution.
That is,
$f_{i}({\bf x}+{\bf e}_{i},t) = f_{i}({\bf x},t) +  (\Omega_{i}^{k})^{c} +
(\Omega_{i}^{k})^{p} + f^{\prime}_{i}({\bf x},t),
$
where $f^{\prime}_{i}$ is chosen so that its stress moment is nonzero,
while conserving mass and momentum. The random stress fluctuations are
uncorrelated in space and time and are sampled from a Gaussian
distribution. The intensity of the random stress represents
the magnitude of local temperature.

We performed critical quenches with $<\psi(x)>=0$, while $<\psi(x)>
\ne 0$ for off-critical quenches. The largest systems simulated were
$1024^2$ in two dimensions and $256^3$ in three dimensions. Although we have
investigated the domain
growth and scaling properties for a variety of lattice sizes and
parameters, we report on the domain growth and dynamical scaling
properties for only one set of parameters. The results obtained with
smaller lattices and different parameters  such as surface tension are
consistent with the data presented here.

 The lattice was initialized with a random distribution of
the different colored fluids. The growth kinetics is characterized
through the order parameter correlation function
 $G(r,t) =
<\psi(r)\psi(0)>-<\psi>^2$, averaged over shells of radius r. The
domain size, $R(t)$, is then defined as the first zero of $G(r,t)$ and
the Fourier transform of $G(r,t)$ is the structure factor $S(k,t)$.

We first discuss the two dimensional results.
The effects of fluctuations on domain growth is shown in Fig.1
 for a binary fluid after a critical quench. The
three cases shown are the growth due to (a) thermally induced
fluctuations as described above (+) (b) fluctuations due to initial
velocity only with average velocity $<u>=0$ and $<u^2> \ne 0$ $(\Box)$ and (c)
both
thermal and initial velocity fluctuations included $(\Diamond)$. If
neither (a) or (b) is present, no patterns are obtained.
Since the energy will decay due to dissipation in the system, for very
long times the systems with initial fluctuations or thermal
fluctuations should behave differently. However, in the current study,
up to the times simulated, the results are consistent with each other
and show that the origin of the fluctuations has little bearing on
domain growth. Thus,
the  previous use [4,5] of an initial
 fluctuation in
velocity only as the driving mechanism for droplet formation  would
seem justified. In all subsequent simulations in this work we have used case
(c).

 In Fig. 2 is shown the domain growth  for a binary fluid after off-critical
quenchs for
($<\psi(x)>=\frac{1}{3}$ $(\Diamond)$
 and $<\psi(x)>=\frac{1}{2}$
(+).
The early time domain growth of  $t^{1/3}$ and the later stage growth of $t^{.4
\pm.02}$   is clearly evident
regardless of the average order parameter.
We  can interpret this as long range diffusion of fluid particles across
growing  domains at early time
giving way to domain growth where hydrodynamic  or inertial effects become
important.
In order to understand the effects of hydrodynamics in the system, the
velocity   $\bf u$ of the particles is set to zero before the
collision step.  For such a system, the convective effect and
dissipative mechanisms will disappear and the only dynamics left is
the diffusive process. In Fig. 3 is shown the domain growth for this
diffusive motion.
The minority phase
forms small droplets at first which then grow very slowly into larger
domains. The growth is very slow and is hampered by transient effects
which slow down the ordering process. A transient regime where the
growth scales as $t^{.23}$, close to $t^{.25}$,
 can be identified and is likely caused by
short range diffusion along the boundary of the domains, as explained by
Mullin\cite{mullin}. This short
range diffusion crosses over at very late times to a $t^{1/3}$
behaviour characteristic of long range diffusion. Though this has been
previously predicted theoretically\cite{lif},  and has been detected using
extrapolation methods, we are not aware of any
off-critical
simulations  in which it has been directly observed.

We expect topological factors to play a significant role in the ordering
process
for three and four component fluid mixtures. We observe that the
domains are rather compact and can eventually be hindered from
growing, similar to the effects of confinement of  a fluid mixture in
a  pore geometry\cite{gru}.  Figure 4  shows the domain growth for three
and four component mixtures in the presence of hydrodynamics. For the
symmetric three $(\Diamond)$ and four (x) component mixtures an early $t^{1/3}$
regime can be
identified which crosses over to a clear $t^{.40\pm.02}$ growth for late times
where the domain morphology is compact. The growth for a three
component system with concentrations (.2,.2,.6) (not shown here)
has similar behaviour.
 Comparison of
Figures 2 and 4 shows that, as expected, a ternary and four component
symmetric fluid mixture behaves like an off-critical binary mixture
with an estimated growth law of $t^{.4\pm.03}$  for late times. The early time
 behaviour for $(\Diamond)$ and (x) shows the $t^{1/3}$ growth characteristic
of long range
diffusion.  Our results clearly indicate that hydrodynamics plays a significant
role in the late
time behaviour, contrary to an earlier MD work\cite{lar}
 which we suspect was not
run  long enough. A Langevin model simulation of Farrell and Valls\cite{far1}
on a 100 by 100 grid
for an off-critical binary quench predicts $t^{.35 \pm .03}$. This
result has been interpreted as showing $t^{1/3}$ behaviour. However,
the unpublished work by Wu {\em et al.}\cite{wu} with this method indicates an
exponent
very close to results reported here.  It appears that large lattice
sizes and long time simulations are necessary to obtain correct growth kinetics
by this method.
 The reason for the similar behaviour of three and four component
fluids as the off critical binary fluid may be understood in terms
of the number of contact points of different phases. The physics appears to
be dominated by the two phase line contacts rather than the finite
number of three or four phase point contacts.

The three dimensional growth is shown in Fig.5 for $256^3$ for a
critical $(\Diamond)$ and
off-critical (+)  quenched binary fluid with $< \psi(x) >= .5$.
Both cases show similar behaviour. Finite size effects have not allowed us
to run simulation long eough to estimate the exponents for later stage growth.
However, if we calculate an effective exponent\cite{huse}, $n_{eff}$, by
considering the domain sizes at time $M t$ and $t$, where M is an integer,
then the extropolated behaviour for large domain is $ R(t) \sim
t^{0.96\pm0.05}$
(see the insert of Fig. 5).
The  results are in agreement with
Siggia's prediction\cite{siggia}.

Experimental studies has been inconclusive in the estimates of growth
exponents.
 The  measurement of an off-critical quench of a
simple acid-water mixture gave a growth exponent
between .32 and .35 \cite{wong},
whereas a   growth rate of $\sim .5$ was observed in  a
very careful study of an off-critical quenched block
copolymer\cite{bat}.
Transient effects and the lack of error analysis  in the
previous studies
 makes direct comparison with experiments difficult.
Thus we can only  stress the need for more  experiments on
off-critical binary
quenches  and
multicomponent fluids.

In conclusion, we have used a lattice Boltzmann model with
fluctuations in the fluid and interface,
 to study the
kinetics of domain growth in two and three dimensions for an off-critical
quenched binary fluid and three and four component fluid
mixtures. We find that the domain growth scales as $t^{.4\pm.03}$ in two
dimensions, indicating that hydrodynamics is relevant to the kinetics of phase
separation as seen in critical quench. Our three dimension results are in
agreement with
phenomenological theory\cite{siggia}.

We thank F. Alexander,  G. D. Doolen, D. W. Grunau,  A. Lapedes and Y. Wu
for useful discussions. The work was supported by the U.S. Department of Energy
at Los Alamos National Laboratory. Numerical simulations were carried out using
the
Connection-Machine 5 at the Advanced Computing Laboratory at the Los Alamos
National Laboratory.

* Department of Applied Mathematics, University of Western Ontario,
London, Ontanio, Canada N6A 5B7

\vspace{0.2in}
\noindent {\bf Figure Captions}
\begin{description}

\item{Figure 1.} The effects of fluctuations on domain growth for a binary
fluid on a $512^2$ lattice after a critical quench. The three cases represent
growth due to (a) thermally induced fluctuations via the stress tensor
(+) (b) fluctuations due to initial velocity perturbation  only
$(\Box)$ and (c) both thermal and initial velocity fluctuations
$(\Diamond)$.

\item{Figure 2.} Domain growth for a binary fluid after an off-critical quench
for $<\psi(x)>=\frac{1}{3}$ $(\Diamond)$ and $<\psi(x)>=\frac{1}{2}$
(+). The straight lines represent fits to $<\psi(x)>=\frac{1}{3}$ of
$\sim t^{1/3}$, showing the early time diffusive growth, and
$\sim t^{.4}$ for the later stage growth where inertial effects are
important.

\item{Figure 3.} Growth for a binary fluid after an off critical quench
for $<\psi(x)>=\frac{1}{3}$ without hydrodynamics. The transient behaviour
due to
interfacial diffusion ($\sim t^{0.23}$) eventually leads to the expected long
range behaviour.

\item{Figure 4.} Growth for a symmetric 3-component system
$(\Diamond)$ $(1/3,1/3,1/3)$ and  a symmetric 4-component mixture ($\times$)
$(1/4,1/4,1/4,1/4)$.

\item{Figure  5.} Three dimensional domain growth $R(t)$ on a
$256^3$ lattice for a critically quenched $(\Diamond)$ and off-critically
quenched ($< \psi(x) >= .5$ (+)) binary fluid, showing similar growth dynamics.
The extropolated growth exponent, $n_eff$
versus $1/R(t)$ as seen in the insert, gives $\sim t^{0.96}$, close to expected
growth law $t^{1}$.

\end{description}

\end{document}